\documentclass{IEEEtran}
\IEEEoverridecommandlockouts
\usepackage{cite}
\usepackage{amsmath,amssymb,amsfonts}
\usepackage{algorithmic}
\usepackage{graphicx}
\usepackage{textcomp}
\usepackage{xcolor}

\def\BibTeX{{\rm B\kern-.05em{\sc i\kern-.025em b}\kern-.08em
    T\kern-.1667em\lower.7ex\hbox{E}\kern-.125emX}}
\begin{document}

\title{Modular Drive Architecture for Software-defined Vehicles Enabled by Power-packet-based\\Sensorless Control}

\author{Shiu~Mochiyama and Rikuto~Kawasome%
\thanks{This work has been submitted to the IEEE for possible publication.}%
\thanks{This work was partially supported by JSPS KAKENHI 24K17262 and Kansai Research Foundation for Technology Promotion.}%
\thanks{The authors are with Kyoto University, Kyoto-daigaku Katsura, Kyoto 615-8510 Japan. Correspondence to: s-mochiyama@dove.kuee.kyoto-u.ac.jp}}

\maketitle

\begin{abstract}
The transition toward software-defined vehicles requires standardization and modularization of hardware decoupled from software, along with centralized electrical/electronic architectures. 
While electrified drive units, such as integrated in-wheel drives, are expected to realize the hardware standardization and unprecedented flexibility in vehicle design, their implementation remains constrained by complex signal wiring between the module and the vehicle body and by control units decentralized across them. 
This paper proposes a modular drive architecture that achieves complete hardware-software separation by leveraging the power packet dispatching system. We introduce a sensorless control method that estimates motor internal states, specifically winding current and rotor angle, solely from physical quantities measured on the vehicle side. 
This completely eliminates the need for physical sensors in the drive module, reducing it to a passive actuator governed by the vehicle-side power system via a standardized packet protocol. The proposed architecture significantly reduces wiring complexity and centralizes control logic, advancing fully standardized, plug-and-play platforms for next-generation electrified mobility. 
\end{abstract}


\section{Introduction}
The Software-Defined Vehicle (SDV) concept is driving a paradigm shift in automotive development, primarily through the separation of hardware and software\cite{Stumpfle.etal-2025}. Decoupling these layers frees system functions from hardware constraints, providing design flexibility to realize various functions via software updates on a single physical platform.

Electrification makes SDVs practically viable. Unlike internal combustion engine vehicles, electric vehicles consist of fewer, simpler components, facilitating hardware standardization. An ultimate example is the modularized drive\cite{wangIntegratedModularMotor2015}, which integrates motors and inverters into a single module. Placing these modules at the vehicle's four corners frees up the central chassis, enabling a standardized \textit{skateboard platform}\cite{Adrian-Chernoff-Skateboard-platform,whiteheadInwheelMotorsRoll2018}. This allows a single platform to serve various mobility applications, significantly reducing development time and costs. Such standardization promotes the mass adoption of compact, cost-effective electrified mobility for urban transport\cite{Grausam.etal-2022}, potentially contributing to carbon neutrality with concepts like Vehicle-to-Grid (V2G)\cite{chenEnergyInformationManagement2020}.

To fully leverage this standardized hardware, a fundamental renewal of the onboard electrical/electronic (E/E) architecture is essential. The industry is rapidly shifting toward a zonal architecture, integrating vehicle control by physical layout rather than functional divisions\cite{Khamis.Goswami-2025,Lim.etal-2025}. Ideally, this architecture consolidates intelligence into a central controller, treating terminal modules as pure actuators. This drastically simplifies wiring harnesses and enables centralized software management.

However, conventional drive systems still exhibit strong hardware-software coupling. To precisely control the modularized drives, dedicated sensors and local control units are typically installed within the module. This \textit{intelligence-at-the-edge} architecture requires complex communication harnesses to the central controller. Furthermore, functional updates demand synchronous firmware adaptations for each module. This strong dependency hinders the flexibility of a true zonal architecture, where adding or modifying hardware should not necessitate software redesign.

This paper addresses the above challenge by introducing a sensorless control approach enabled by power packetization. The power packet dispatching system (PPDS) physically integrates power and information by discretizing power transfer in the time domain and attaching information tags, enabling the strict trace of each power packet flow via its tag\cite{Takuno.etal-2010}. Furthermore, the network between the sources and loads consists of unified hardware called power packet routers\cite{Takahashi.etal-2015} and their operations are software-reconfigurable\cite{Inagaki.etal-2021}. (A technical explanation of how these functions are implemented is provided in Section~\ref{sect:packet}.) Leveraging the complete traceability of the PPDS, we eliminate sensors from the drive module by establishing a method to estimate internal motor states (winding current and rotational angle) solely from physical quantities measurable on the vehicle-side power packet transmission line.

The primary significance of this study lies in reducing the drive module to a passive component that operates via commands in a standardized \textit{power packet} protocol from a central, software-defined power management system. This approach enables the following three key innovations. 
First, complex signal wiring is eliminated, resulting in an extremely simple structure connected solely by power packet lines (harness reduction).
Second, by centralizing control logic and state estimation in the vehicle-side software, functional enhancements are completed entirely through central updates without modifying terminal hardware.
Third, motors with different specifications can be easily replaced in a plug-and-play manner by merely updating the central software, provided their interface is standardized to power packets.

\section{Power Packet Dispatching System} \label{sect:packet}
This section first provides an overview of the configuration and operating principles of the power packet dispatching system (PPDS) based on existing literature. 
Furthermore, we define the vehicle-side power system to which the proposed drive module is connected and discuss its software-definability in detail. 

\subsection{Power Packet\cite{Takuno.etal-2010}} 
\begin{figure}
    \centering
    \includegraphics[width=\linewidth]{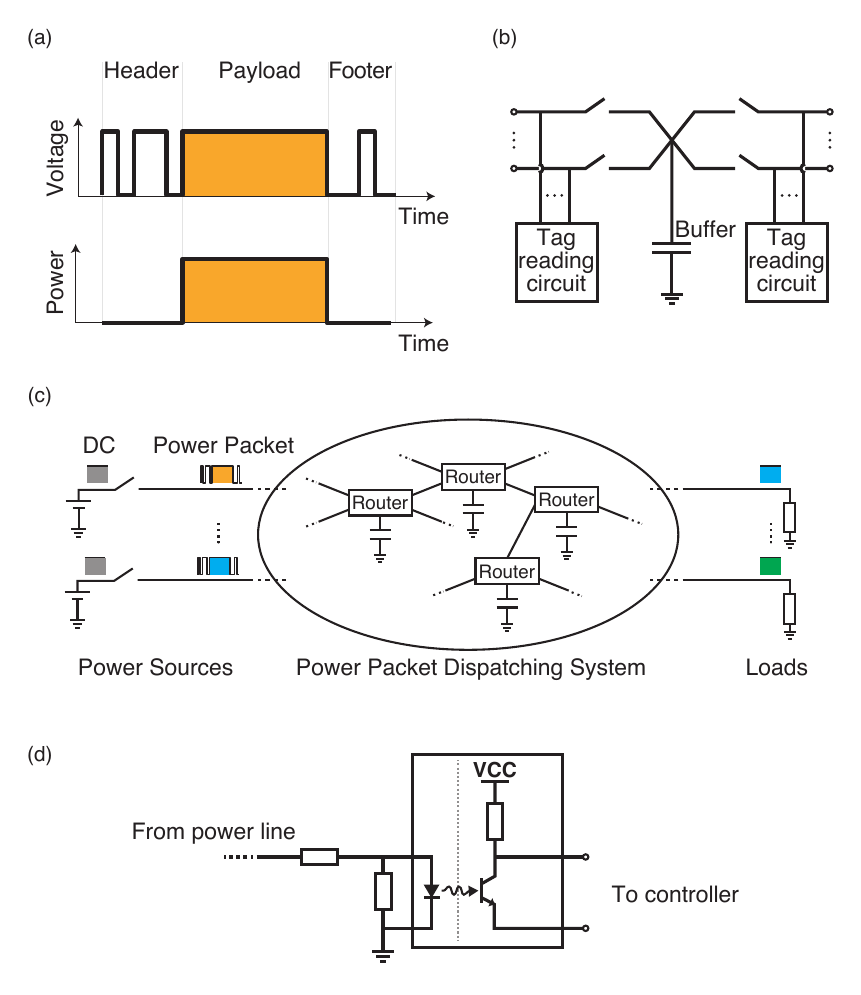}
    \caption{Power packet dispatching system (PPDS). (a) Configuration of a power packet. (b) Configuration of a power packet router. (c) Routers' network to constitute a PPDS. (d) Configuration of a tag reading circuit. Some elements are adapted from the authors' prior publication\cite{Mochiyama.Hikihara-2023}.}
    \label{fig:ppds}
\end{figure}

A power packet consists of an information tag, represented by a voltage signal, and a power pulse called a payload. 
Fig.~\ref{fig:ppds}~(a) illustrates the configuration of a power packet. 
The information tag is divided into a header and a footer. 
The header is attached immediately before the payload and carries information regarding the transmission path and instructions for load control. 
The footer is attached after the payload to indicate the end of the packet. 

\subsection{Power Packet Routing} 
Power packets are transmitted from power sources to loads via a network of power packet routers (hereafter simply referred to as routers)\cite{Takahashi.etal-2015}.  
Fig.~\ref{fig:ppds}~(b) shows the basic configuration of a router, and Fig.~\ref{fig:ppds}~(c) illustrates an example of a PPDS comprising a network of the routers. 
The primary functions of a router are threefold: (1) reading the information tag of a transmitted packet, (2) determining whether to receive the packet based on the acquired information, and (3) forwarding the packet to the subsequent router. 

First, tag reading is performed by a reading circuit that serves as an isolated interface between the power line and the router's controller\cite{mochiyamaPowerPacketDispatching2021}. 
Fig.~\ref{fig:ppds}~(d) shows an example configuration of the reading circuit. 
Using a voltage divider and an isolation IC, such as a photocoupler, the information tag signals are safely transmitted to the controller. 

Next, upon deciding to receive the packet based on the information tag, the controller turns on the switch corresponding to the incoming port to receive the payload power. 
The received payload power is temporarily stored in a capacitor. 
During this period, the reading circuit continuously monitors the voltage envelope of the packet; when the end information from the footer is detected, the switch is turned off, terminating the reception. 
If the router determines not to receive the packet, the switch remains turned off. 

Finally, the power temporarily stored in the capacitor is re-packetized by controlling the switch corresponding to the appropriate output port and then transmitted to the next router. 
The information tag is attached by turning the switches on and off according to the signal logic. 
Of course, if the next destination is a load, the tag attachment procedure is omitted, and only the payload is transmitted to the load\cite{mochiyamaPowerPacketDispatching2021}. 

In the PPDS, unlike conventional continuous-flow control based on PWM, power transmission is controlled by modulating a digital sequence composed of packetized power units. 
Previous studies have proposed a power modulation method inspired by signal quantization\cite{Takahashi.etal-2016a}. 
In essence, this is modulation based on the transmission density of power packets, namely Packet Density Modulation (PDM), and it has been experimentally shown to achieve performance comparable to PWM\cite{Mochiyama.Hikihara-2019a}. 

\subsection{Vehicle-side Power System Using Power Packet} 
If a router is viewed as a device that determines the spatial input-output relationship of power packets, its operation is governed by local rules configured within each router and information shared from adjacent routers via tags. 
Given that the PPDS treats power at the router's input/output as time-discrete and quantized digital sequences, this input-output relationship can be seen as analogous to logic operations in information theory. 
Previous studies have experimentally demonstrated that the router's input-output relationship can be defined by logic operations, which is redefinable on the same hardware\cite{Inagaki.etal-2021}, and that power computation can be performed by interconnecting such routers\cite{Mochiyama.Hikihara-2023}. 
Consequently, the input-output relationship of the entire network and the spatio-temporal distribution of power within the network can be modified by updating the operational rules without altering the hardware. 
This confirms that the PPDS serves as a software-definable power system, as discussed in the Introduction. 

\subsection{Comparison with Similar Approaches}
Another approach to constructing software-definable power systems is the proposal of Software-Defined Power Electronics (SDPE) \cite{Zhou.Preindl-2024}. 
It realizes software-defined power converters by integrating modular elements (Power Electronics Building Blocks \cite{Ericsen.Tucker-1998}) with hierarchical control. 
This allows various converter topologies for different applications to be implemented using common hardware through software updates, with applications reported in onboard power system interfacing with the grid\cite{Zhou.Preindl-2023}. 

While power systems constructed via SDPE follow the conventional power electronics concept of building systems by connecting power converters in series or parallel via a common bus, the PPDS discretizes power transport into packets and performs energy management based on autonomous routing by a router network. 
The packetization of power physically isolates the power flow between different sources and loads. 
Therefore, bus voltage instability caused by constant power loads such as motors, which is a known issue in such multi-converter systems\cite{Emadi.Ehsani-2001,Emadi.etal-2006}, does not occur in principle (See also \cite{Mochiyama.etal-2025} for detailed discussion on this point). 
The PPDS offers advantages in scalability regarding component changes or additions, as there is no need to redesign the entire energy management system when modifying designs for specific applications or when exchanging or adding loads. 

Based on the above discussion, the objective of this paper is to extend the inherent software-definability of the PPDS to motor drive modules. 
In other words, we aim to realize a drive module that achieves both the unique plug-and-play concept, enabled only with the PPDS, and the software-definability by merely rewriting the vehicle's main controller, which governs the router operation of the vehicle-side network. 
To this end, in the following sections, we establish a power-packet-based modular drive architecture, which completely eliminates the sensors inside the module and communication harnesses between the module and the main (vehicle-side) controller. 

\section{Modular Drive Architecture} \label{sect:method}

\subsection{Hardware Configuration}
\begin{figure*}
    \centering
    \includegraphics[width=0.85\linewidth]{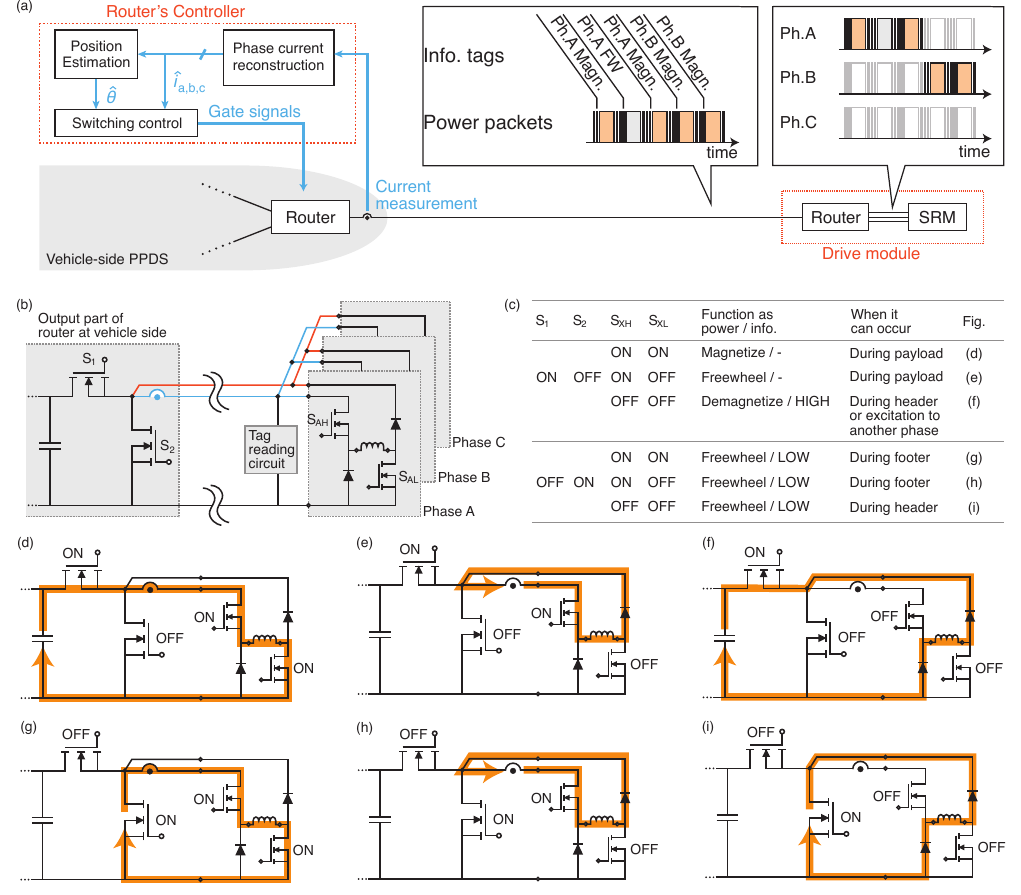}
    \caption{Proposed modular drive architecture. (a) Overall configuration. The main components of the architecture are the consolidated intelligence in the vehicle-side controller, enabled by vehicle-side current measurement and the drive module operating passively according to the information tag. (b) Circuit configuration for the power packet transmission. (c) Switching table and functions of each switching state. (d)--(i) Current path in each switching state. The corresponding switching states are indicated in the rightmost column of (c). }
    \label{fig:module}
\end{figure*}
The proposed modular drive architecture is illustrated in Fig.~\ref{fig:module}~(a).
This configuration comprises two routers, one at the vehicle side (vehicle-side router) and another at the drive module side (module-side router) and a switched reluctance motor (SRM). 
Fig.~\ref{fig:module}~(b) depicts the detailed circuit configuration. 
The vehicle-side router has the same configuration as a standard power router in a general PPDS. 
The module-side router features an input port that interfaces with the vehicle-side power system, and output ports corresponding to the excitation phases.
Comparing this router with a standard PPDS router reveals several circuit modifications. 
First, the output stage switches are replaced with a modified type of asymmetric half-bridge converters proposed in \cite{ganOnlineSensorlessPosition2016}. 
Second, the input stage switches, which let in or block an incoming power packet in a normal PPDS router\cite{Takahashi.etal-2015}, are omitted. This is a reasonable simplification because the same function can be performed by switching the output stage only.

Fig.~\ref{fig:module}~(c) describes the switching tables and the functions of the circuit in each switching state. 
During a header and a footer, the vehicle-side router expresses the logic HIGH or LOW by turning on S$_1$ and off S$_2$ or on S$_2$ and off S$_1$, respectively. 
During a payload, S$_1$ is kept on and S$_2$ off. 
In this period, the module-side router operates in one of three modes for each of the three excitation phases: magnetizing mode (both upper and lower switches are ON; Fig.~\ref{fig:module}~(d)), freewheeling mode (only the upper switch is ON; Fig.~\ref{fig:module}~(e)), or demagnetizing mode (both switches are OFF; Fig.~\ref{fig:module}~(f)). 
In addition to these modes, three types of freewheeling path occur during a header or footer transmission (corresponding to Fig.~\ref{fig:module}~(g)--(i)), which will be explained later in the description of the protocol of a power packet transmission. 

Before detailing the proposed architecture, we explain the rationale for adopting an SRM as the traction motor in the drive module. 
First, SRMs are being actively studied as a viable option for the electrification of commercial vehicles\cite{Beheshti.etal-2026}. 
Although permanent magnet synchronous motors (PMSMs) are currently the most common selection due to their high torque density and efficiency, the use of rare-earth permanent magnets poses economic and geopolitical risks. 
SRMs are expected to offer a rare-earth-free alternative to this issue. 
Specifically, the motor's inherent low cost makes it a best suited option for the mass introduction of new mini-mobility solutions\cite{Grausam.etal-2022}. 
Second, SRMs are excited by a DC current due to its unipolar drive requirement. 
This inherently aligns with the packetized power supply discussed in this paper. 

\subsection{Protocol for Power Packet Transfer}
The information tag of the power packet transmitted from the vehicle-side power system contains signals indicating the start and end of the packet, as well as a phase excitation command to be executed by the module-side router.
The packet start information is set as a 1-bit HIGH signal, and the end signal as a 4-bit LOW signal.
The phase excitation command specifies either the magnetizing mode or the freewheeling mode for one of the three excitation phases and demagnetization mode for the others.
The combination of the excitation phase and its mode is mapped to a 3-bit signal.

In this paper, we assume that only one excitation phase is in the magnetizing mode at any given time.
This is because the current reconstruction method described later relies on the time-division multiplexing characteristics of power packets, and concurrent excitation of multiple phases disrupts this principle.
While this does not pose a problem in the subsequent discussions, some motor drive methods involve simultaneous excitation of multiple phases.
Even in such cases, setting the duration of a power packet sufficiently short and allocating power packets to multiple phases over a given proportion of time slots realize pseudo-simultaneous supply in the sense of time-average, as demonstrated in previous studies on multiple load control using power packets\cite{Mochiyama.Hikihara-2019a}.

The vehicle-side power system determines the aforementioned phase excitation commands based on the motor's phase currents and rotor's angle, which are estimated using the method described later.
This paper adopts the hysteresis current control method, which is common in SRM drives, to determine the command in the following procedure. 
First, after detecting the phase switching timing, the phase to be excited is shifted in the rotational direction, and the header and payload are transmitted with the command indicating that the corresponding phase is set to the magnetizing mode.
Subsequently, when the estimated current of the phase being magnetized reaches the upper limit of the hysteresis band, a footer is transmitted.
Following this, the header indicating that the phase is set to the freewheeling mode and an empty payload are transmitted.
When the estimated current of the phase reaches the lower limit of the hysteresis band, a footer is transmitted.
Then, the header indicating that the phase is set back to the magnetizing mode and a payload are transmitted.
This process is repeated until the phase switching timing is detected. Once detected, the cycle returns to the beginning.

The module-side router controls the ON/OFF states of the output stage switches based on the signals from the reading circuit.
Upon detecting the start of a packet, it selects switching states for the three phases that correspond to the operation mode indicated by the 3-bit command signal, namely Fig.~\ref{fig:module}~(d), (e), or (f). 
These gate signals are maintained until the end of the power packet is detected by reading a footer signal. 
For a footer transmission, before the last bit of a footer is read, the switching state corresponds to Fig.~\ref{fig:module}~(g) or (h), depending on the command for this packet. 
Then a header of the next packet is transmitted, during which the switching state corresponds to Fig.~\ref{fig:module}~(f) or (i).

\subsection{Current Reconstruction}
The reconstruction of motor phase currents in the vehicle-side power system is performed by combining the command attachment to the headers with the current measurement during the subsequent payloads at the vehicle-side sensor.
Specifically, when phase $X = \{A, B, C\}$ is in the magnetizing or freewheeling mode, the current measured during the corresponding payload is estimated as the current of phase X, and the estimated currents of the other phases are set to zero.

In this method, current estimation cannot be performed in some period of the operation. 
First, in the demagnetization mode, namely while the current is decreasing to zero at the end of the phase's excitation period, the current measured in the vehicle-side power system does not match the motor phase current. 
This can be confirmed in Fig.~\ref{fig:module}~(f). 
Such an unmeasurable period does not affect the proposed system's operation because the phase in the demagnetization mode is not the target of the current control. 
Second, the same situation also occurs during the header transmission, namely while the circuit configuration alters between those indicated by Fig.~\ref{fig:module}~(f) and (i).  
This unmeasurable period also does not affect the operation because the command in the header is decided at the beginning of the packet based on the current measurement during the last packet. 

It should be noted that this method assumes that the router on the drive module side correctly follows the commands.
This assumption is valid since the vehicle-side router and the module-side router are directly connected and there are no other routers in between and therefore packet collisions and subsequent transmission failures cannot occur in principle. 
Nonetheless, failures in reading the tag disrupt the operation; thus, ensuring robustness against noise is a crucial technical requirement for realizing this method. 
One approach is to introduce redundancy for error correction in the header communication protocol, although we did not do it just for simplicity in the experimental implementation. 

\subsection{Turn-off Angle Detection}
The key to estimating the phase switching timing is using the results of the current reconstruction alongside the relationship between current and rotational position derived from the motor's electromechanical characteristics.
Numerous such methods have been proposed in the literature\cite{Xiao.etal-2022}, and among them, we explain the specific procedure by taking the method proposed in \cite{ganOnlineSensorlessPosition2016} as an example. 
The same method will also be used for the experimental verification in the next section.

\begin{figure}
    \centering
    \includegraphics[width=0.8\linewidth]{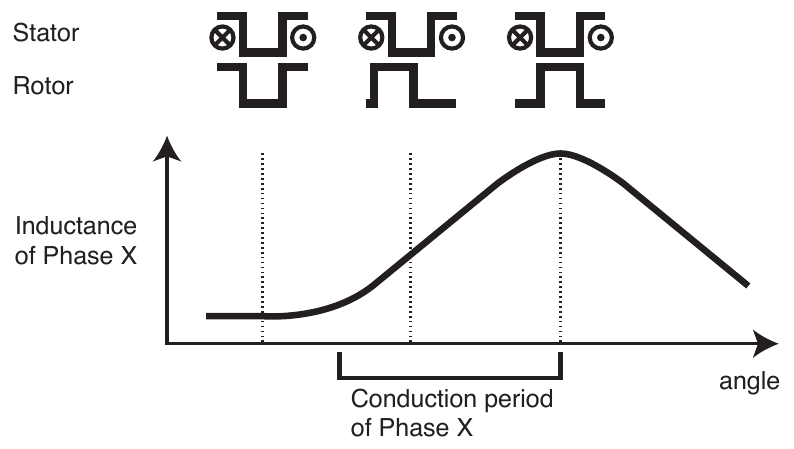}
    \caption{Inductance versus angle characteristics.}
    \label{fig:inductance}
\end{figure}

First of all, the winding inductance depends both on current due to the characteristics of the magnetic material and on rotor angle due to saliency.
Here, since the hysteresis control tries to maintain the current around a constant value $I_\mathrm{ref}$, we assume the current is constant.
Then, the inductance with respect to the rotational angle is as shown in Fig.~\ref{fig:inductance}.
The inductance reaches its maximum when the stator and rotor poles are aligned, and its minimum when they are unaligned (located between adjacent aligned states).
Therefore, we set the timing of phase switching when the inductance takes its maximum value to drive the motor without negative torque generation. 

Next, we consider the detection of the maximum inductance. 
During the conduction period of a phase $X=\{A,B,C\}$, namely from the beginning of the switch to the phase until the switching to the next phase, the phase $X$ current is governed by
\begin{equation} \label{eq:de}
    L(i_X,\theta)\frac{\mathrm{d}i_X}{\mathrm{d}t} = -ri_X + e(\omega) + E,
\end{equation}
where $E$ is the voltage applied across the windings, $L(i_X,\theta)$ is the winding inductance, $r$ is the winding resistance, and $e(\omega)$ represents the electromotive force dependent on the rotational speed $\omega$.
During the conduction period of phase $X$, the current repeats to rise to the upper band (magnetizing mode) and fall to the lower band (freewheeling mode) by the hysteresis control. 
Assuming that the current dependence of the inductance and the change in rotational speed are negligible during the conduction period, we introduce the following relationship as an approximation from Eq.~(\ref{eq:de}):
\begin{equation} \label{eq:de_approx}
    L(I_\mathrm{ref},\theta_n) \cdot m_n = -rI_\mathrm{ref} + E',
\end{equation}
where $\theta_n$ and $m_n$ are the angle and current slope when $i=I_\mathrm{ref}$ in the $n$-th magnetizing mode, respectively, and $E'=e(\omega)+E$ is regarded as a constant based on the assumptions.
From Eq~(\ref{eq:de_approx}), for two consecutive magnetizing modes, the following relationship holds:
\begin{equation} \label{eq:final}
    L(I_\mathrm{ref},\theta_n) \cdot m_n = L(I_\mathrm{ref},\theta_{n+1}) \cdot m_{n+1}.
\end{equation}
Eq~(\ref{eq:final}) implies that, under the aforementioned assumptions, relative changes in the inductance value can be detected from relative changes in the current slopes.
Combining this with the preceding discussion on current reconstruction, the phase switching timing can be detected using the current measurement at the vehicle-side router.

\section{Experimental Verification}

\subsection{Setups}
We set up an experimental system consisting of a vehicle-side router and the proposed drive module with an SRM. 
To focus on the verification of the proposed drive module, we assume the appropriate operation of the vehicle-side PPDS and connect a regulated bidirectional power source to the input of the vehicle-side router. 
Specifically, the power source is connected to the left-hand side of the circuit described in Fig.~\ref{fig:module}~(b). 
We employ a 6/4 SRM with permanent magnets installed in the stator yoke to enhance the torque/power density\cite{Kucuk.Nakamura-2020}, which was also used in the preliminary feasibility study of the PPDS-based SRM drive\cite{Mochiyama.Nakamura-2023}. 
The mechanical configuration of the motor gives the target of the phase switching angle of $45^\circ$. 

The controllers of the vehicle-side router and the proposed module are both implemented in the DSP (B-Box RCP 3.0, Imperix Ltd.). 
However, to maintain the concept of a purely passive drive module, we completely separated the codes for the two controllers. 
Namely, the only connection between the two controllers is via the transmission of information tag. 

The target current of hysteresis control is set at $I_\mathrm{ref}=0.8\,\mathrm{A}$ and the hysteresis band width is set at $0.1\,\mathrm{A}$. 
The voltage of the regulated power source is set to $24\,\mathrm{V}$. 

The experimental setup is equipped with an angle sensor attached to the rotor shaft and current sensors attached to the phase windings, both within the drive module, to evaluate the results with the proposed method. 
We emphasize that they are for validation purpose only, and that our aim is to demonstrate that the proposed system controls the SRM without using these sensors. 

To verify the switching angle detection, we adopt an indirect approach instead of directly using the estimated switching timing to determine the drive command. 
Specifically, we determine the command using the actual angle measured by the aforementioned sensor. 
The turn-on and turn-off angles are set at $30^\circ$ and $60^\circ$, respectively. 
Note that this range includes the estimation target of $45^\circ$. 
Alongside the above procedure, the proposed detection algorithm runs in the vehicle-side controller separately. 
Then, we compare the estimation result and the actual angle to confirm the validity of the proposed method. 
This indirect verification avoids potential fatal damage to the testing bench caused by unexpected errors. 

\subsection{Results}
\begin{figure}
\centering
\includegraphics[width=\linewidth]{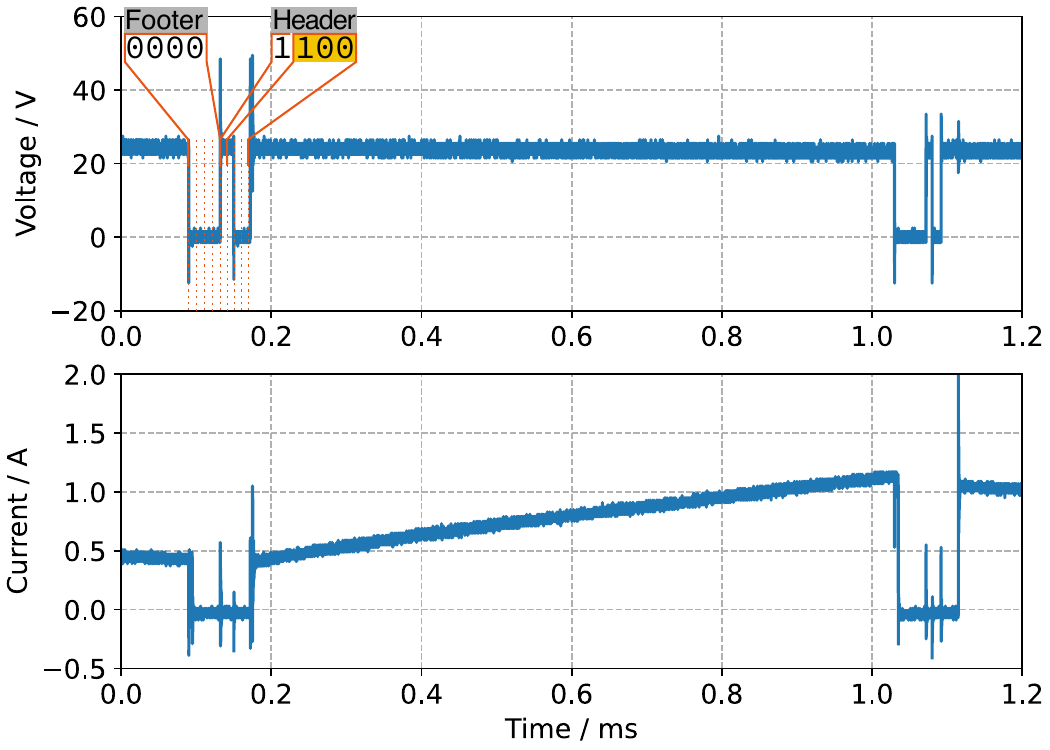}
\caption{Voltage and current waveforms of a power packet. Top: Voltage across the output port of the vehicle-side router. Bottom: Current measured by the sensor placed on the vehicle side. }
\label{fig:result_tag}
\end{figure}

First, we verify the power packet routing based on the information tags. Fig.~\ref{fig:result_tag} shows the voltage and current waveforms of the power packet transmission measured at the output port of the vehicle-side router. As seen in the voltage waveform, the header tag information is 1100, which designates the command "magnetize phase B." The current waveform indicates that the current remains zero during the tag transmission, starts flowing after the tag ends, and returns to zero upon transitioning to the footer period. The beginning of the footer is triggered when the current reaches the predefined threshold of $1.1\,\mathrm{A}$. These observations confirm that the power packet transmission is executed as designed.

\begin{figure}
\centering
\includegraphics[width=\linewidth]{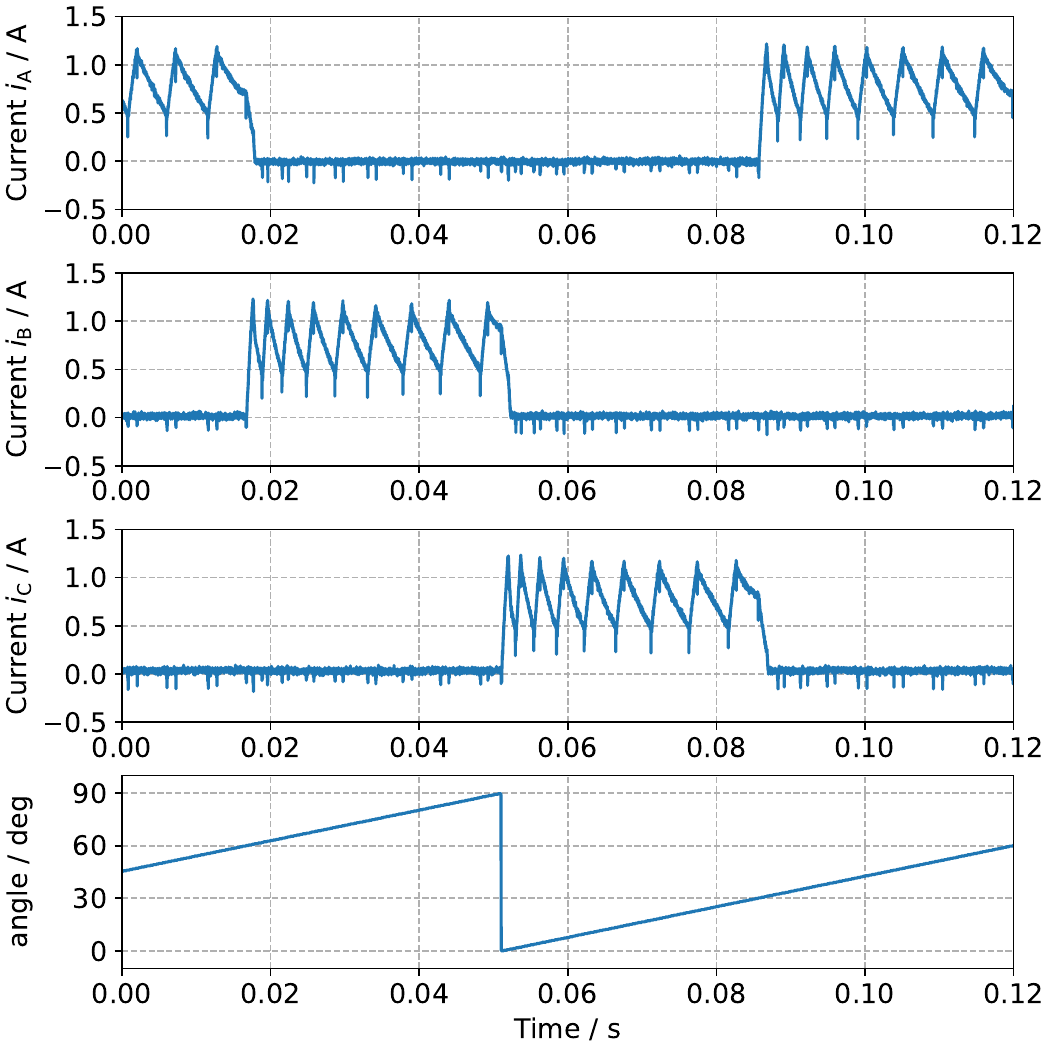}
\caption{Current Waveforms measured at motor phase A (top row), B (2nd row), and C (3rd row) and actual angle trajectory (bottom row). }
\label{fig:phase_current}
\end{figure}

Next, we verify the SRM drive via power packet routing. Fig.~\ref{fig:phase_current} presents the phase-current and rotor's angle waveforms, all measured by the sensor placed inside the module. 
The angle is plotted as a modulo-90 value. 
It can be seen that the current is successfully controlled to stay within the hysteresis band $0.8\pm 0.3\,\mathrm{A}$. 
The small spikes at around the band edges are due to the relatively low gate resistance for the MOSFETs and it is possible to mitigate them by optimizing the gate drives. 
Then looking at the angle trajectory, the motor was driven smoothly with the packetized power supply. 
Here, we can confirm that the excitation phase is switched in accordance with the configured conduction angles, namely $30^\circ$--$60^\circ$ for phase A.

\begin{figure}
\centering
\includegraphics[width=\linewidth]{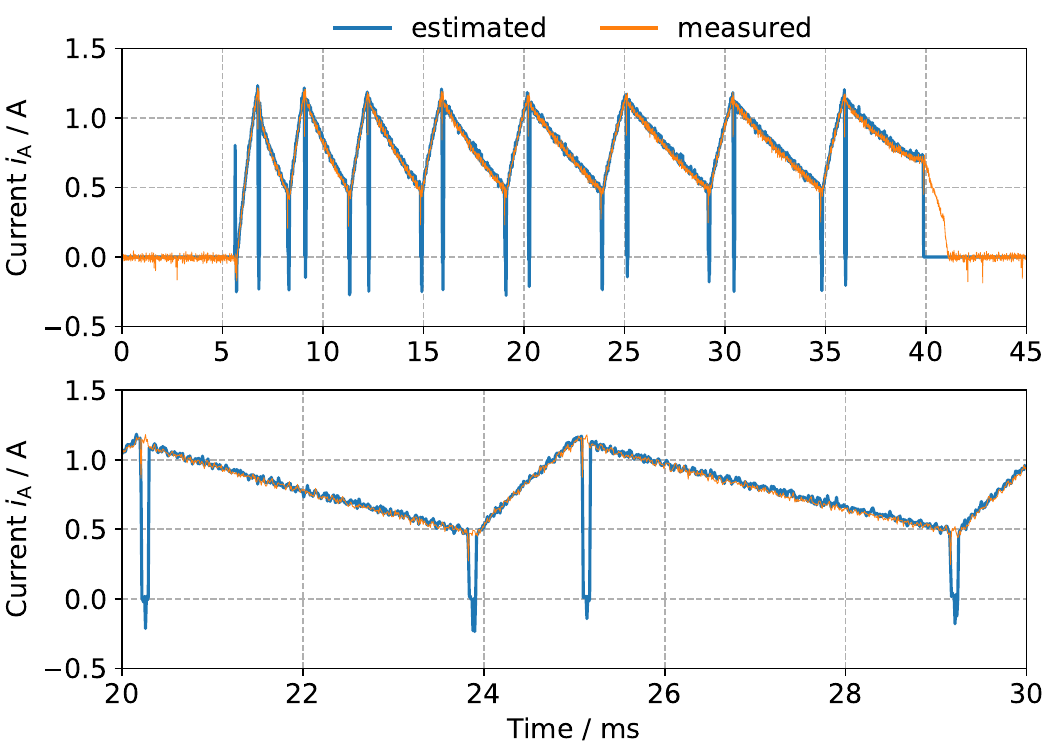}
\caption{Measured and reconstructed current waveforms (top) and its enlarged view (bottom). }
\label{fig:current_reconst}
\end{figure}

We then evaluate the results of the current waveform reconstruction. Fig.~\ref{fig:current_reconst} compares the measured and reconstructed current waveforms for phase A. The figure demonstrates that the actual phase current is accurately reconstructed for most of the operating period. Discrepancies between the measured and reconstructed waveforms only occur during the information tag transmission and the tail current remaining after the switch to phase B. As discussed in Section~\ref{sect:method}, these periods cannot be reconstructed due to the inherent principles of the proposed method; however, this limitation does not affect the operation of the proposed sensorless scheme. 

\begin{figure}
\centering
\includegraphics[width=\linewidth]{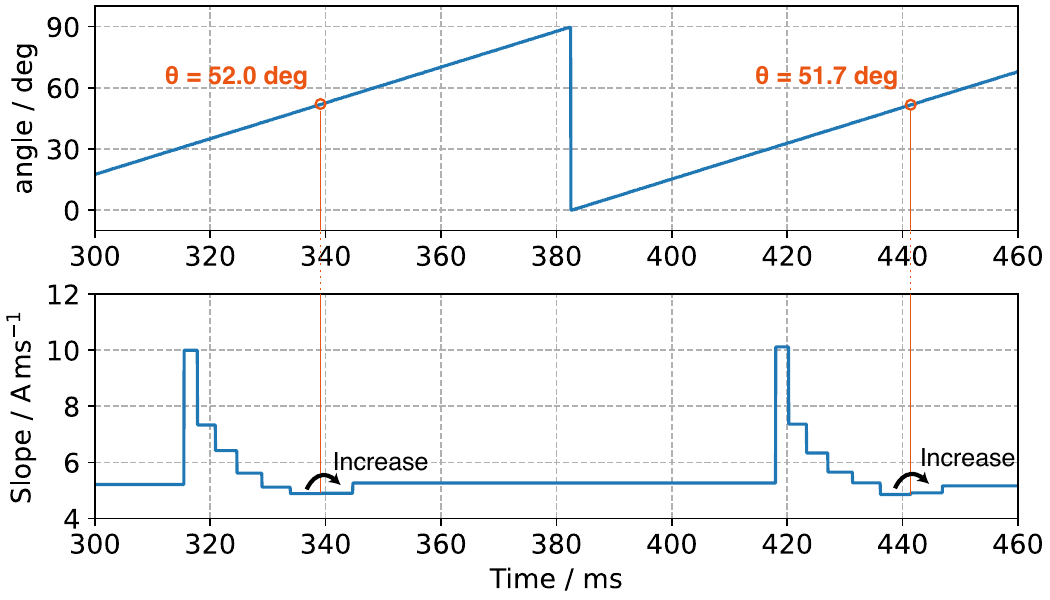}
\caption{Result of turn-off angle detection. Top: Actual angle trajectory with orange circles indicating the angle at the timing of slope increase detection. Bottom: Current slope sequence. }
\label{fig:angle_detect}
\end{figure}

Lastly, we confirm the results of the turn-off angle detection. Fig.~\ref{fig:angle_detect} shows the actual angle trajectory and the current slope during each magnetizing mode of phase A. 
The slope is calculated using the reconstructed current. 
We configured the calculation to retain the most recent value while other phases are excited; thus, only the values within the excitation interval are valid. 
The slope monotonically decreases for a while after switching to phase A and then starts increasing around the target angle. 
Note that, in our experimental setup, the duration of each magnetizing mode is approximately $1\,\mathrm{ms}$ and the detection is made around $1\,\mathrm{ms}$ after the rotor passes through the exact target angle. 
The small delay between the target and detected switching timings can thus be reduced by increasing the switching frequency, specifically by narrowing the hysteresis band or increasing the supply voltage, for example. 
The continuous performance improvements in recent wide-bandgap devices makes such adjustments highly feasible in practical implementation. 
Based on the above discussion, we conclude that the slope change of the reconstructed current can indicate the appropriate timing for phase switching. 

\section{Conclusions}
In this paper, we developed a modular drive architecture as an essential element to realize fully software-defined vehicles (SDVs). By integrating power packetization with sensorless state estimation, this study demonstrated the feasibility of complete hardware-software separation. Experimental results verified that the vehicle-side power system can reconstruct motor phase currents and detect switching timings. The module successfully drove the motor through the purely passive operation of distributing power packets according to their tags, entirely eliminating the need for local controllers and signal harnesses within the drive module.

While this architecture significantly advances standardized platforms for next-generation electrified mobility, future work will focus on expanding the packet protocol to accommodate simultaneous multi-phase excitation for advanced torque control. Furthermore, evaluating and enhancing the robustness of information tag transmission against electromagnetic noise in a setup closer to real-world conditions remains a key step toward practical deployment. 

\section*{Acknowledgment}
The authors express their sincere gratitude to Professor Takashi Hikihara and Professor Taketsune Nakamura of Kyoto University for their invaluable advice. The authors are also deeply grateful to Professor Nakamura for his generous assistance in developing the SRM and its testing environment.

\bibliographystyle{IEEEtran}
\bibliography{references}

\end{document}